\documentclass[twocolumn,showpacs,amsmath,aps]{revtex4}
\usepackage{graphicx,color}
\usepackage{CJK}
\usepackage{bm}
\usepackage[hypertex]{hyperref}
\usepackage{float}

\newcommand{\be}{\begin{equation}}
\newcommand{\ee}{\end{equation}}
\newcommand{\bea}{\begin{eqnarray}}
\newcommand{\eea}{\end{eqnarray}}
\newcommand{\bsube}{\begin{subequations}}
\newcommand{\esube}{\end{subequations}}

\newcommand{\Eq}[1]{Eq.\,(\ref{#1})}

\newcommand{\dg}{\dagger}
\newcommand{\la}{\langle}
\newcommand{\ra}{\rangle}

\newcommand{\nl}{\nonumber \\}

\newcommand{\beq}{\begin{equation}}
\newcommand{\eeq}{\end{equation}}
\newcommand{\beqn}{\begin{eqnarray}}
\newcommand{\eeqn}{\end{eqnarray}}
\newcommand{\bsub}{\begin{subequations}}
\newcommand{\esub}{\end{subequations}}

\begin{document}
\begin{CJK*}{GBK}{Song}

\title{Theory for frequent measurements of spontaneous emissions
        in non-Markovian environment: beyond Lorentzian spectrum}

\author{Luting Xu}
\email{xuluting@tju.edu.cn}
\affiliation{Center for Joint Quantum Studies,
School of Science, Tianjin University, Tianjin 300072, China}

\author{Xin-Qi Li}
\email{xinqi.li@tju.edu.cn}
\affiliation{Center for Joint Quantum Studies,
School of Science, Tianjin University, Tianjin 300072, China}

\begin{abstract}

The measurement-result-conditioned evolution
of a system (e.g. an atom) with spontaneous emissions of photons
is well described by the quantum trajectory (QT) theory.
In this work we generalize the associated QT theory from infinitely wide bandwidth
Markovian environment to the case of finite bandwidth non-Markovian environment.
In particular, we generalize the treatment for arbitrary spectrum,
which is not restricted by the specific Lorentzian case.
We rigorously prove a general existence of a perfect scaling behavior
jointly defined by the bandwidth of environment
and the time interval between successive photon detections.
For a couple of examples, we obtain analytic results
to facilitate QT simulations based on the Monte-Carlo algorithm.
For the case where analytical result is not available,
numerical scheme is proposed for practical simulations.
\end{abstract}

\maketitle

\section{introduction}

In quantum mechanics, the theory of quantum measurement is based on
the Copenhagen's postulate of wave-function collapse
conditioned on a specific outcome of measurement of an observable.
Later generalization of the measurement theory,
largely related to some indirect coupling
and/or taking into account the realistic
microscopic degrees of freedom of the apparatus,

has been more generally formulated as some POVM forms \cite{NC00}.
In quantum optics, for continuous measurement,
the theory has also been formulated as quantum trajectory (QT)
and has found potential applications \cite{WM09,Jac14,Dali92,WM93}.
In recent years, the QT theory has been applied as well
to experiments in superconducting solid-state circuits
\cite{Pala10,DiCa13,Hof11,Mar11,Dev13,Sid13,Sid12,DiCa12,Dev13a}.

Actually, the existing QT theory is based on quantum measurements
in Markovian (wide bandwidth) environments.
A generalization of the QT theory for continuous
measurements performed in non-Markovian (finite bandwidth)
environment was recently proposed \cite{Xu16}, where the non-Markovian environment was modeled
by using a Lorentzian spectral density function (SDF)
with bandwidth ($\Lambda$),
and perfect ``scaling" property was found between
the spectral bandwidth and and the measurement time interval ($\tau$),
in terms of the scaling variable $x=\Lambda\tau$ \cite{SG14}.
This generalization bridges the gap
between the existing QT theory \cite{WM09,Jac14} and the Zeno effect,
by rendering them as two extremes corresponding to
$x\to\infty$ and $x\to 0$, respectively.

Therefore, a question on how we extend the continuous measurement
theory to more general non-Markovian environment beyond the Lorentzian SDF, along with the question regarding the possibility of generally finding and proving
the $x=\Lambda\tau$ scaling behavior.
In the present work, we provide an investigation for this problem.
The work is organized as follows.
In Sec.\ II we present a numerical scheme to calculate
the null-results conditioned state evolution,
where the numerical accuracy is examined
by comparing with the analytic results of the Lorentzian SDF,
before carrying out the numerical results in Sec.\ III
for three non-Lorentzian examples.
Here perfect scaling behavior is first observed
and then rigorously proved for general case.
In Sec.\ IV we outline the Monte-Carlo algorithm
and display the simulation results of quantum trajectories.
Finally, we summarize the work with discussions in Sec.\ V.

\section{Formulation and Method}

Let us consider a two-level atom coupled to
electromagnetic vacuum (environment),
which is described by the Hamiltonian
\begin{align}
H= \frac{\Delta}{2}\sigma_z
+\sum_r\left(b^{\dg}_{r}b_r+\frac{1}{2}\right)\omega_r
+\sum_{r} \left(V_{r} b^{\dg}_{r}\sigma^{-}
     + {\rm H.c.} \right) \, .
\end{align}
Throughout this work we set $\hbar=1$.
Here we introduce: the two-level energy difference
$\Delta=E_e-E_g$,
the atomic operators
$\sigma_z=|e\ra \la e|-|g\ra \la g|$, $\sigma^-=|g\ra \la e|$,
and $\sigma^+=|e\ra \la g|$. $V_{r}$ is the radiative
coupling of the atom to the environment.
Let us consider the evolution of the entire system,
starting with an initial state
$|\Psi(0)\rangle=(\alpha_0|e\rangle+\beta_0|g\rangle)
\otimes |{\rm vac}\ra$,
where $|{\rm vac}\ra$ stands for the environmental vacuum
with no photon. Under the influence of coupling,
the entire state at time $t$ can be expressed as
\bea\label{WF-1}
|\Psi(t)\rangle &=& \alpha(t)|e\rangle\otimes |{\rm vac}\ra
+ \sum_r c_r(t)|g\rangle\otimes |1_r;0;{\cdots}\ra  \nl
&& + \, \beta_0|g\rangle\otimes |{\rm vac}\ra  \,,
\eea
where $|1_r;0;{\cdots}\ra$ describes the environment
with a photon excitation in the state ``$r$"
and no excitations of other states.
The coefficients have initial conditions of
$\alpha(0)=\alpha_0$ and $c_r(0)=0$.

Substituting Eq.~\eqref{WF-1}
into the Schr\"odinger equation,
$i\partial_t |\Psi(t)\rangle =H|\Psi(t)\rangle$
and performing the Laplace transformation,
$\tilde f(\omega) =\int_0^\infty f(t)\exp (i \omega t)dt$,
we obtain the following system of algebraic equations:
\begin{subequations}
\label{eqs}
\begin{align}
&(\omega-E_e)\tilde \alpha (\omega)-\sum_r V_{r}\tilde c_r(\omega)
=i\alpha_0 \,,
\label{eqs1}\\
&[\omega-(E_g+\omega_r)]\tilde c_r(\omega)-V^*_{r}\tilde \alpha(\omega)=0 \,.
\label{eqs2}
\end{align}
\end{subequations}
The r.h.s. of the first equation reflects the initial condition.
Substituting $\tilde{c}_r(\omega)$ from Eq.~(\ref{eqs2})
into Eq.~(\ref{eqs1}), we obtain
\begin{align}
&(\omega-E_{e})\tilde \alpha(\omega)
-{\cal F}(\omega) \tilde \alpha(\omega )=i\alpha_0 \, .
\label{a2}
\end{align}
In this result, we have introduced
\begin{equation}  \label{Fw}
\mathcal{F}(\omega)=
\int\frac{D(\omega_r)}{\omega-(E_g+\omega_r)}\, d\omega_r  \,,
\end{equation}
where the spectral density function (SDF)
was defined as usual as
\bea
D(\omega_r)=\sum_{r'}|V_{r'}|^2 \delta(\omega_r-\omega_{r'}) \,.
\eea

Rather than the wide-band limit required for the ``Markovian'' reservoir,
in this work we consider a finite-band spectrum.
For Lorentzian SDF, as shown in Ref.\cite{Xu16}, we can first solve
\Eq{a2} in frequency domain, then obtain the analytic
solution of $\alpha(t)$ by means of inverse-Laplace transformation.
However, for arbitrary SDF $D(\omega_r)$,
this strategy does not work.
Instead, we can solve \Eq{a2} for $\alpha(t)$ numerically in time domain.
For this purpose, an inverse-Laplace transformation to \Eq{a2} yields
\begin{align}
\dot{\alpha}(t)=-iE_e\,\alpha(t)-i\int_{0}^{t} du \, F(u) \, \alpha(t-u) \,,
\label{a-t}
\end{align}
where the kernel function in the integral in time domain
is the inverse Laplace transformation of ${\cal F}(\omega)$, which reads
\bea\label{F-t}
F(u) = -i\, \int d\omega_r D(\omega_r)\, e^{-i(\omega_r+E_g)u} \,.
\eea
In obtaining this result, we have used the following identity
\bea
\int^{\infty}_{-\infty} \frac{d\omega}{2\pi}\,
\frac{e^{-i\omega u}} {\omega-(\omega_r+E_g)}
 = -i\,e^{-i(\omega_r+E_g)u} \, .
\eea
It would be desirable to eliminate the energy ($E_e$) caused phase factor
and the initial condition $\alpha(t=0)=\alpha_0$.
Let us introduce the {\it decay factor} of the excited state $a(t)$,
via $\alpha(t)=\alpha_0 e^{-iE_et} a(t)$,
and introduce as well $\tilde{F}(u) = F(u)e^{iE_e u}$.
We have
\begin{equation}\label{til-alph-t}
\dot{a}(t)=-i\int_{0}^{t} du \, \tilde{F}(u) \, a(t-u) \,.
\end{equation}
In practice, for a given SDF $D(\omega_r)$,
one can first carry out $\tilde{F}(u)$ in advance,
then numerically integrate \Eq{til-alph-t} to obtain $a(t)$
using a discretized algorithm as follows
\begin{align}\label{til-alph-t-2}
a(Ndt)=a[(N-1)dt]
-i\,dt \left(
\sum_{j=1}^{N} \tilde{F}(jdt) a[(N-j)dt]\, dt \right)
\end{align}
Here $N\,dt=t$, with $dt$ a discretized increment of time interval.

\subsection{Lorentzian Spectrum: Analytic Solution}

Let us consider first the Lorentzian SDF which allows analytic solution.
The Lorentzian SDF is assumed as
\begin{align}
D(\omega_r)= D_0\Lambda^2/[ (\omega_r-\omega_0)^2+\Lambda^2]\, ,
\label{lor}
\end{align}
where $\omega_0$ is the spectral center, $D_0$ the spectral height
and $\Lambda$ the spectral width.
Substituting this SDF into \Eq{Fw}, we obtain
\begin{align}
\mathcal{F}(\omega )
={\Lambda\Gamma/2 \over (\omega-\omega_0-E_g)+i\Lambda},
~~{\rm where}~~
\Gamma=2\pi D_0 \,.
\end{align}
Based on this result, one can solve \Eq{a2} first for
$\tilde \alpha(\omega )$ in the frequnecy domain,
then perform an inverse Laplace transform,
$\alpha(t)=\int_{-\infty}^\infty \tilde \alpha(\omega)
e^{-i\omega t}d\omega /(2\pi)$.
Taking into account the convention
$\alpha(t)=a(t)\alpha_0 e^{-iE_e t}$, one obtains
the decay factor \cite{SG14}
\begin{align}
a(t)={1\over A_+^{}-A_-^{}}(A_+^{}e^{-A_-^{}t}-A_-^{}e^{-A_+^{}t}) \,,
\label{proj0}
\end{align}
where $A_{\pm}^{}=[\Lambda -iE\pm
\sqrt{(\Lambda -iE)^2-2\Gamma\Lambda}]/2$,
and the energy offset $E=(E_e-E_g)-\omega_0$.

Let us now consider to introduce frequent projective measurements
of photon in the reservoir, with time intervals $\tau$.
We may have two possible results: a photon registered by the detector;
or no photon detected, i.e., a {\it null} result of measurement.
For the first case, the atom jumps to its ground state.
For the case of null result, the atom state would also have a change
by excluding the second term with one photon in the reservoir from \Eq{WF-1}.
At this moment, we are interested in the case of $n$ successive null results
of measurements. Conditioned on this, the atom state at $t=n\tau$ should be
\bea\label{null-Psi}
|\Psi_A (t)\ra = \left([a(\tau)]^n \alpha_0 |e\ra + \beta_0 |g\ra \right)/{\cal N} \,,
\eea
where ${\cal N}$ denotes a normalization factor.

We may further consider the limit of ``continuous" measurements,
$n\to\infty$ by taking the measurement time interval
$\tau\to 0$ and keeping $t=n\tau$ fixed.
Supposing to increase the bandwidth $\Lambda$
so that the variable $x=\Lambda\tau$ remains constant,
we can prove a ``scaling" property
that the final state becomes a function of $x$ only.
To be a little bit more general,
we also assume the energy offset $E=c\Lambda$
(in usual treatment $c=0$).
{\bf
After simple manipulations}, we arrive to \cite{SG14}
\begin{align}\label{bar-a}
\bar a(t)=[a(\tau)]^n=\exp\left\{
-\left[\frac{1}{\kappa}-( 1-e^{-\kappa x})\frac{1}{\kappa^2x}
\right]\frac{\Gamma t}{2}  \right\}   \,,
\end{align}
where $\kappa=1-ic$.
Elegantly, this {\it effective decay factor}
reveals an explicit scaling property
in the $x=\Lambda \tau$-variable.

\subsection{Accuracy Examination }

For non-Lorentzian SDF, it may not be possible to
obtain analytic solution as above for the Lorentzian spectrum.
However, instead, one can carry out numerical results.
Before applying the numerical method to several examples,
we would like first to examine it by comparison
with the analytic solution under Lorentzian spectrum.
The key quantity for numerical computation
is the kernel function $\tilde{F}(u)$,
which should be obtained in advance.
For the Lorentzian SDF,
this kernel function can be easily obtained as
\bea
\tilde{F}(u)&=& -i\, \int d\omega_r D(\omega_r)\, e^{-i(\omega_r+E_g-E_e)u} \nl
&=& -i\,(\Gamma\Lambda/2) e^{i (E_e-E_g-\omega_0) u}e^{-\Lambda u}  \,.
\eea
With this, one can solve for $a(t)$ from \Eq{til-alph-t},
i.e., numerically from the iterative expression \Eq{til-alph-t-2}.
In Fig.\ 1(a), for atom initially in $\alpha_0|e\ra+\beta_0|g\ra$ and later
subject to the influence of spontaneous emission (coupling to environment),
we show the decay probability of the component $|e\ra$, $|a(t)|^2$.
We compare the results based on the numerical \Eq{til-alph-t-2}
and from the analytic solution \Eq{proj0}, respectively,
for spectral bandwidths $\Lambda=100\Gamma$, $10\Gamma$, $5\Gamma$ and $\Gamma$.
In Fig.\ 1(b), we further compare the results of the decay probability
of $|e\ra$, conditioned on null-results of
frequent measurements with time interval $\tau$.
In the numerical calculation, we adopt $\Lambda=5\Gamma$ and
a couple of $\tau$ resulting in thus
$x=\Lambda\tau=2$, 0.2 and 0.02.
We compare the results against the analytic solution of \Eq{bar-a}
which are obtained under the limits $\Lambda\to\infty$ and $\tau\to 0$.
The full agreement of the results shown in Fig.\ 1(a) and (b)
demonstrate that the numerical method proposed above
is efficient and precise enough,
which can be safely applied to the non-Lorentzian spectrum
where analytic solution may not be available.

\begin{figure}[H]
  \centering
  \includegraphics[scale=0.45]{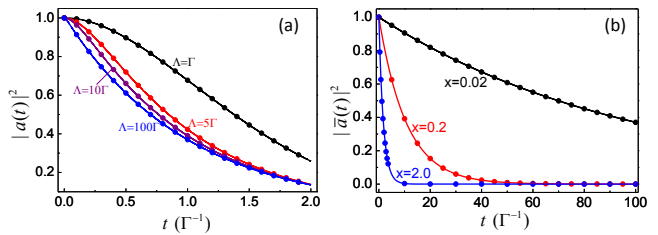}
  \caption{
Accuracy examination of the numerical method via comparison
with the analytic solution under Lorentzian spectrum.
(a)
Decay probability of the excited state $|e\ra$,
under the influence of spontaneous emission (coupling to environment)
but without intermediate interruptions of frequent null-result measurements.
The dots are from the numerical \Eq{til-alph-t-2}
while the lines from the analytic solution \Eq{proj0}.
(b)
Decay probability of $|e\ra$ conditioned on
the frequent null-result measurements.
The dots are from numerical calculation based on \Eq{til-alph-t-2}
and $\bar{a}(t)|_{t=n\tau}=[a(\tau)]^n$,
by adopting $\Lambda=5\Gamma$ and a couple of $\tau$ so that
$x=\Lambda\tau=2$, 0.2 and 0.02.
The lines display the analytic solution of \Eq{bar-a}
under the limits $\Lambda\to\infty$ and $\tau\to 0$.
We also assumed $E_e-E_g=\omega_0$.  }
\end{figure}

\section{Numerical Results and General Scaling Behavior}

In this section we apply \Eq{til-alph-t} or (\ref{til-alph-t-2})
to several non-Lorentzian examples.
We will consider in particular the null-results conditioned
evolution under continuous measurement,
which is a key ingredient for the construction of quantum trajectories.
That is, assuming $\alpha_0=1$
(initially the atom on the excited state $|e\ra$),
we will compute the survival probability
$P_e(t)=|\bar{a}(t)|^2$, where $t=n\tau$ and $\bar{a}(t)=[a(\tau)]^n$.
Numerically, we obtain $a(\tau)$ from \Eq{til-alph-t}.

Also, based on the numerical results, we will first explore
the existence of scaling behavior, then rigorously prove it
for arbitrary SDF.

\subsection{Three Models}

Let us consider the following three models of non-Lorentzian SDF.

These models may approximately describe possible real systems
but here they are largely used for mathematical purposes.
As we will see soon, these specific models allow us to obtain
analytic expressions of $\tilde{F}(u)$
which can guide us to find a property necessarily needed
to prove the interesting scaling behavior.
However, we will prove subsequently
that the scaling behavior holds for arbitrary SDF,
not depending on any specific forms such as the models we assume here.

The first example is a Gaussian SDF given by
\begin{equation}
D(\omega_r)=D_0e^{-\frac{(\omega_r-\omega_0)^2}{2\Lambda^2}}  \,.
\end{equation}
Accordingly, we obtain the kernel function in \Eq{til-alph-t} as
\begin{align}
\tilde{F}(u)= -i\, \frac{\Gamma\Lambda}{\sqrt{2\pi}}
e^{i (E_e-E_g-\omega_0)u}e^{-\Lambda^2 u^2} \,,
\end{align}
where $\Gamma=2\pi D_0$.

The second example we will consider is a constant SDF
with finite bandwidth, given by
\begin{equation}
D(\omega_r)=\left\{
\begin{aligned}
&D_0,\,\,&|\omega_r-\omega_0|\leq\Lambda/2\\
&0,\,\,&|\omega_r-\omega_0|>\Lambda/2
\end{aligned}
\right.  \,.
\end{equation}
For this model, the corresponding kernel function reads
\begin{align}
\tilde{F}(u)= -i\, \frac{\Gamma}{\pi u}
e^{i (E_e-E_g-\omega_0)u}\sin(\Lambda u/2) \,.
\end{align}
Here we also defined $\Gamma=2\pi D_0$.

The third example is a double-Lorentzian SDF
\begin{equation}
D(\omega_r)= \sum_{j=1,2}
\frac{D_0\Lambda^2}{\Lambda^2+[\omega_r-\omega_0
+ (-1)^j \omega_1]^2}   \,.
\end{equation}
Similarly, we obtain the kernel function as
\begin{align}\label{2Lorentz}
\tilde{F}(u)= -i\, \Gamma\Lambda
e^{i (E_e-E_g-\omega_0) u}e^{-\Lambda u}\cos(\omega_1 u) \,.
\end{align}
We defined as well  $\Gamma=2\pi D_0$.


\begin{figure}[H]
  \centering
  \includegraphics[scale=0.6]{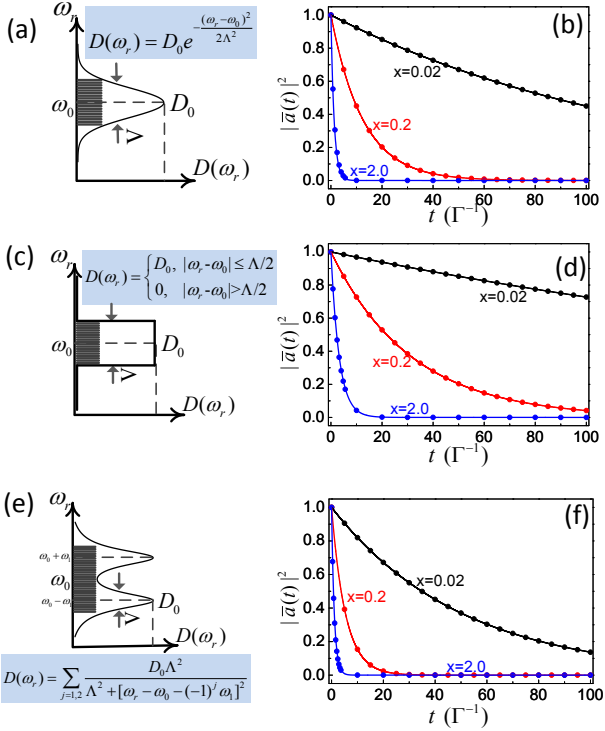}\\
  \caption{
Frequent null-results conditioned decay of $|e\ra$
under coupling to environment with non-Lorentzian spectrum.
Three examples of SDF are schematically shown in (a), (c) and (e), while
the corresponding results are displayed, respectively, in (b), (d) and (f).
For each given $x$, different $\Lambda$ (and accordingly $\tau$)
are chosen, e.g., $\Lambda=100\Gamma$ (lines) and $5\Gamma$ (dots)
as shown here. }
\end{figure}

\subsection{Numerical Results}

In Fig.\ 2 we plot the results for the three non-Lorentzian examples.
That is, for each SDF, we show the null-results conditioned decay
probability of $|e\ra$, i.e., $|\bar{a}(t=n\tau)|^2=|a^n(\tau)|^2$ .
We plot results for three parameters $x=2.0$, 0.2 and 0.02.
We observe that the decay is slowed down as we reduce $x$,
as a consequence of the Zeno effect.
This is because by noting that $x=\Lambda\tau$, for a fixed $\Lambda$,
smaller $x$ simply means
smaller $\tau$ (i.e. more frequent measurements).
More interesting observation is that,
for a given $x$ but quite different $\Lambda$
(e.g., $\Lambda=100\Gamma$ and $5\Gamma$ as we compare in Fig.\ 2),
the null-results conditioned evolution is identical
under varying $\Lambda$ and accordingly $\tau$,
as shown in Fig.\ 2 by the perfect coincidence of the dots and the lines.
This clearly reveals a remarkable `scaling' behavior,
with $x=\Lambda\tau$ the scaling variable.

Qualitatively, we may understand the scaling behavior as follows,
by means of the time-energy uncertainty principle.
According to the uncertainty principle,
the measurements with time interval $\tau$
will disturb the atomic level of $|e\ra$
by an energy fluctuation of $\tau^{-1}$.
Then, for more frequent measurements (smaller $\tau$),
if we expand as well the spectral width $\Lambda$ of the reservoir
to keep $x=\Lambda\tau$ unchanged
(see the examples schematically plotted in Fig.\ 2),
the atomic decay (spontaneous emission)
is seemingly not to be affected by the stronger energy fluctuations
($\sim \tau^{-1}$) with respect to the reservoir spectrum.
However, it is still somehow surprising
that the scaling behavior governed by $x=\Lambda\tau$
is so precise, as shown in Fig.\ 2
by the perfect coincidence between the dots and curves.
It would be of great interest to investigate further,
more quantitatively, the underlying reason.

\subsection{Proof of Scaling Behavior}

In our previous work \cite{Xu16,SG14}, restricted in the Lorentzian SDF,
we have proven an exact scaling property via obtaining the analytic
expression \Eq{bar-a}, under the `continuous' limit
$\tau\to 0$ (meanwhile making $\Lambda\to\infty$).
For non-Lorentzian case, however, the analytic solution of $a(t)$
such as \Eq{proj0} is not available.
Below we present a different proving method,
which is applicable to broader cases.

To fulfill such a proof, a key observation is that,
for all the examples illustrated above,
the rescaled $\tilde{F}(u)/\Lambda$ is
a function of the {\it joint} parameter $x=\Lambda u$.
We may formally denote it as
\bea\label{gx}
\tilde{F}(u)/\Lambda=g(x) \,.
\eea

Note that here we have involved the consideration
$E=c\Lambda$, while $E\equiv E_e-E_g-\omega_0$ is
the offset of the transition energy from the spectral center.
Also, in \Eq{2Lorentz} for the double-Lorentzian SDF,
the locations of the peak centers are proportional to the peak width,
i.e., $\omega_1\propto\Lambda$.

Starting with \Eq{til-alph-t}, let us consider the evolution
over $(0,\tau)$.
For the limit $\tau\to 0$, we can replace $a(\tau-u)$
in the integrand by $a(\tau)$, yielding
\begin{equation}\label{at-mkv}
\dot{a}(\tau)=-i\int_{0}^{\tau} du \, \tilde{F}(u) \, a(\tau) \,.
\end{equation}
The physical meaning of this procedure is that,
in short time limit, the non-Markovian memory effect is not relevant.
For this {\it time-local} differential equation,
the solution simply reads $a(\tau)=e^{Q(\tau)}$.
We further evaluate $Q(\tau)$ as follows:
\bea
Q(\tau)&=&-i\int_0^{\tau} d u' \int_0^{u'} du^{\prime\prime}
\tilde{F}(u^{\prime\prime})  \nl
&=& -\,\frac{i \tau}{x}\int_0^x dx^\prime \int_0^{x^\prime}
dx^{\prime\prime} g(x^{\prime\prime}) \,.
\eea
Here we have introduced
$x=\Lambda\tau$, $x^{\prime}=\Lambda u^{\prime}$
and $x^{\prime\prime}=\Lambda u^{\prime\prime}$.
Based on this result, we can easily obtain the
successive null-results conditioned evolution as
\bea\label{abar-2}
\bar{a}(t)=a^n(\tau)=e^{nQ(\tau)}=e^{-\gamma(x)\, t/2} \,.
\eea
Here we have used $t=n\tau$, and the effective decay rate is given by
\bea\label{gamma-x}
\gamma(x)= i\, \frac{2}{x}\int_0^x dx^\prime\int_0^{x^\prime}
dx^{\prime\prime} g(x^{\prime\prime}) \,.
\eea
This result fully proves the {\it scaling property}
of the null-results conditioned evolution.
The mathematical rigorousness is based on the key structure of \Eq{gx},
i.e., $\tilde{F}(u)/\Lambda=g(\Lambda u)$.

Then, the most interesting question is that this structure
can be valid in general for arbitrary SDF?
We now extend our consideration to general case.
Actually, for an arbitrary form of SDF $D(\omega_r)$, we can rewrite it as
$D(\omega_r)= \tilde{D}(\omega_r-\omega_0)$, with
$\omega_0$ close to or simply equal to the atomic transition energy.
Then, as schematically shown in Fig.\ 3,
let us consider {\it a series of deformation of this SDF}
i.e., $\tilde{D}(\omega_r-\omega_0)$, by varying its ``width" $\Lambda$
in accordance with the change of the detection time interval $\tau$,
in order to keep $x=\Lambda\tau$ fixed.
Notice that, for an arbitrary form of SDF, it may not necessarily have
a natural width parameter $\Lambda$ in its functional form.
In order to describe {\it the whole class of deformed curves of the SDF}
as shown in Fig.\ 3 (and explained above),
we can simply introduce a width parameter $\Lambda$ to the SDF, as
$\tilde{D}(\frac{\omega_r-\omega_0}{\Lambda})$.
That is, via varying $\Lambda$ in this function, we can obtain
all the deformed SDFs as schematically shown in Fig.\ 3.
Now let us re-denote this class of SDF by $D(\omega_r,\Lambda)$, i.e.,
$D(\omega_r,\Lambda)\equiv \tilde{D}(\frac{\omega_r-\omega_0}{\Lambda})$.
Obviously, this general consideration renders
the SDF with a ``natural" width parameter
(as numerically demonstrated in the previous subsections)
as special examples of this general form.
Then, we manipulate the proving as follows:
\bea\label{gx-prove}
&& \tilde{F}(u)= -i\, \int d \omega_r \,
D(\omega_r,\Lambda) \, e^{-i(\omega_r+E_g-E_e)u} \nl
&& =
-i\, \int d\omega_r \, \tilde{D}(\frac{\omega_r-\omega_0}{\Lambda})\,
\, e^{-i[(\omega_r-\omega_0)-(E_e-E_g-\omega_0)] u}  \nl
&& = -i\Lambda \int d\omega \tilde{D}(\omega) e^{-i(\omega-c)(\Lambda u)} \nl
&& \equiv \Lambda \, g(\Lambda u) \,.
\eea
Here, as done previously, we have assumed $E_e-E_g-\omega_0\equiv E=c\Lambda$.
In this way we proved the general structure
of $\tilde{F}(u)/\Lambda=g(\Lambda u)$,
which can guarantee the scaling behavior of the null-result conditioned evolution
as proved through Eqs.\ (\ref{at-mkv})-(\ref{gamma-x}).

\begin{figure}[!htbp]
  \centering
  \includegraphics[width=7.5cm]{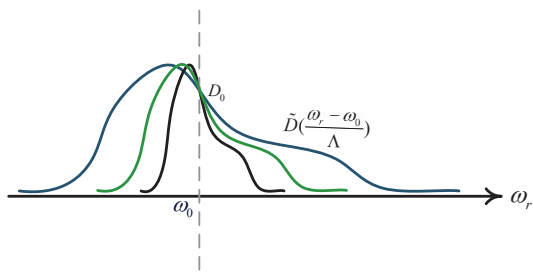}
  \caption{
Deformation of the spectral-density-function (SDF)
via introducing a changeable width parameter $\Lambda$,
in order to keep $x=\Lambda\tau$ fixed
when changing the detection time interval $\tau$.  }
\end{figure}

We would like to mention that,
instead of the short-time-limit treatment as \Eq{at-mkv},
an alternative treatment following Ref.\ \cite{Kur00}
can result in the $x=\Lambda\tau$ scaling behavior as well,
{\it provided that the property proved by \Eq{gx-prove} is valid as it is}.
Moreover, as proved in Appendix A,
both limiting treatments give essentially the same result of \Eq{gamma-x}.
In Ref.\ \cite{Kur00}, rather than the scaling behavior concerned here,
the main interest was concentrated on the Zeno and anti-Zeno effects,
by considering only the change/decrease of $\tau$ but with $\Lambda$ unchanged.
The anti-Zeno effect occurs for some special SDF when speeding the
successive measurements (reducing the interval $\tau$),
manifested as acceleration of decay in certain intermediate range of $\tau$.
As an interesting addition, in the above, we generally proved that
even for the anti-Zeno SDF analyzed in Ref.\ \cite{Kur00},
the decay rate will be exactly the same
--not influenced by speeding the measurements--
if we alter $\Lambda$ as well to keep $x=\Lambda\tau$ unchanged.

We may further carry out the explicit expressions of $\gamma(x)$
for the examples we illustrated.
Before that, we first examine the Lorentzan SDF \Eq{lor}.
Straightforwardly, from \Eq{gamma-x}, we obtain
\begin{eqnarray}
\gamma(x)= \Gamma \left[\frac{1}{\kappa}
-( 1-e^{-\kappa x})\frac{1}{\kappa^2x} \,
\right]
\end{eqnarray}
where $\kappa=1-ic$ and $c$ is introduced from $E=c\Lambda$.
Desirably, based on this different method
(i.e. not solving $a(t)$ to get \Eq{proj0} previously),
we obtain the same result of \Eq{bar-a}.

Now, for the Gaussian SDF shown in Fig.\ 2(a), we obtain
\begin{eqnarray}\label{gmx-1}
\gamma(x)=\Gamma\left[{\rm erf}\left(\frac{x}{\sqrt{2}}\right)+\frac{2}
{\sqrt{2\pi}x}\left(e^{-\frac{x^2}{2}}-1\right)\right] \,,
\end{eqnarray}
where ${\rm erf(x)}=\int_0^xe^{-x^2}dx$ is the Error function.
For simplicity, here and for the other two non-Lorentzian examples,
we assume $E_e-E_g=\omega_0$ (thus $c=0$).
For nonzero $c$, analytic expressions are also available
but more complicated in form.
The results of the other two examples are, respectively,
for the rectangular SDF (Fig.\ 2 (c))
\begin{eqnarray}\label{gmx-2}
\gamma(x)=\frac{2\Gamma}{\pi}\left[{\rm Si}(\frac{x}{2})+\frac{2}{x}
\cos\frac{x}{2}-\frac{2}{x}\right] \,,
\end{eqnarray}
where ${\rm Si}(x)=\int_0^x\frac{\sin x}{x}dx$;
and for the Double-Lorentzian SDF (Fig.\ 2(e))
\begin{align}\label{gmx-3}
\gamma(x)=\Gamma\left(1-\frac{e^{-x}\sin x}{x} \right) \,.
\end{align}
In addition to $c=0$, in this last example we also assumed
$b=1$ with $b$ defined from $\omega_1= b\Lambda$.

\section{Quantum trajectories}

We now turn to constructing a practical scheme
for quantum trajectory (QT) simulation,
associated with continuous measurement (photon detection)
in a non-Markovian environment
and beyond Lorentzian SDF for the atom-environment coupling.
It is well known that the existing QT theory is associated with
continuous measurements in wide-band-limit Markovian environment
\cite{WM09,Jac14,Dali92,WM93}.
From the fundamental theoretical viewpoint, the existing QT theory
has an imperfection that its prediction differs from the
quantum Zeno effect, as explained in detail in Ref.\ \cite{Xu16}.
It would thus be very desirable to develop a unified description
to quantitatively bridge the gap between the QT theory and the Zeno physics.
This problem has been addressed in our recent studies \cite{Xu16,SG14},
where the consideration was restricted
within the finite-bandwidth Lorentzian SDF.
Using Lorentzian SDF, the available {\it analytic solution} allows
not only proving the scaling behavior,
but also constructing the QT approach
based on the {\it effective} emission rate
associated with frequent detection of photons.

Below we extend the quantum trajectory study beyond the Lorentzian SDF,
by applying \Eq{til-alph-t} or (\ref{til-alph-t-2}),
or more efficiently, \Eq{abar-2} to calculate the null-results conditioned
evolution under continuous measurement,
which is a key ingredient for the construction of quantum trajectories.  
In order to simulate the quantum trajectories,
let us consider also to introduce optical drive to the atom.
Conditioned on the results of measurement, one can then
construct a Monte-Carlo wave function approach,
closely following the line of the standard quantum trajectory theory
for measurements in Markovian environment \cite{Dali92,WM93,WM09,Jac14}.

Specifically, let us consider the evolution over
the time interval $(t,t+\Delta t)$.
To construct an {\it efficient} theory for the successive
photon detections with shorter time intervals $\tau$,
one can utilize the accumulated result over $\Delta t=n\tau$
to perform a one-step update for the atom state.
This longer time duration, $\Delta t$,
is roughly determined from the criterion that during $\Delta t$
there is {\it at most} one photon registered in the detector
\cite{Dali92,WM93,WM09,Jac14}.
Therefore, during $\Delta t$, there will be two possible outcomes:
a photon registered in the detectors ($\Delta N_c=1$),
or no photon registered ($\Delta N_c=0$).
In the former case, we simply update the atom state
by a {\it jump} action; while for the latter result the atom
takes an {\it effective} smooth (but non-unitary) evolution.

\begin{figure}[h]
  \centering
  \includegraphics[scale=0.45]{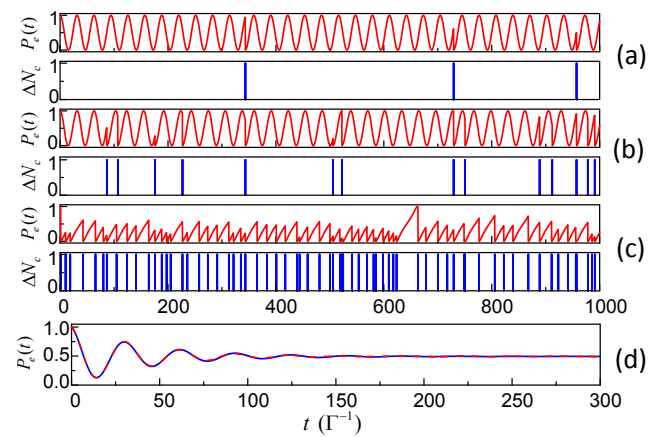}\\
  \caption{
Quantum trajectories, taking the rectangular SDF
shown in Fig.\ 2(c) as an example,
are displayed for $x=0.02$, 0.2 and 2 in (a), (b) and (c), respectively.
The events of photon emissions are also shown by plotting
$\Delta N_c=1$ resulted from the Monte-Carlo simulation.
In (d), for $x=0.2$,
the ensemble-average over 5000 trajectories is compared
against the result from the Lindblad master equation
with photon emission rate $\gamma_{\rm eff}$.
For all the results here, the Rabi parameter $\Omega$ in the driving
Hamiltonian $\Omega\sigma_x$ is adopted as $\Omega=\Gamma$. }
\end{figure}

To perform Monte-Carlo simulations, during $\Delta t$,
the probability with a photon registered in the detectors
is $p^{(n)}_{1}(\Delta t)=|\alpha(t)|^2 \gamma_{\rm eff}\Delta t$.
Here we denote the {\it effective} emission rate
under frequent detections by $\gamma_{\rm eff}$,
which is given by
\bea\label{rate-2}
\gamma_{\rm eff}= [1-|\bar{a}(\Delta t)|^2]/\Delta t \,.
\eea
For small $\Delta t$ and for the form of \Eq{abar-2}, we simply have
$\gamma_{\rm eff}={\rm Re}\gamma(x)$, as given by Eqs.\ (\ref{gmx-1}),
(\ref{gmx-2}) and (\ref{gmx-3}), respectively,
for the examples we illustrated.

In practical simulations,
generate a random number $\epsilon$ between 0 and 1.
If $\epsilon < p^{(n)}_{1}(\Delta t)$,
which corresponds to the probability of having
a photon registered in detectors ($\Delta N_c=1$),
we update the state by a ``jump" action.
Otherwise, the atom experiences a smooth evolution.
Including together the evolution caused by the optical drive,
we can update the atom state in a compact way formally expressed as
\bea\label{M01}
|\Psi_A(t+\Delta t)\ra = {\cal U}(\Delta t)
{\cal M}_{1,0}(\Delta t) |\Psi_A(t)\ra
\, /\parallel\bullet \parallel  \,,
\eea
where $\parallel\bullet \parallel$ denotes the normalization factor.
${\cal U}(\Delta t)$ describes the unitary evolution
owing to the optical drive, while ${\cal M}_{1,0}(\Delta t)$
are the Krause operators in the POVM formalism which read, respectively,
${\cal M}_{1}(\Delta t)=\sigma^-$ for $\Delta N_c=1$,
and ${\cal M}_{0}(\Delta t)=diag\{\bar{a}(\Delta t),1 \}$ for $\Delta N_c=0$.
Note also that the above form of ${\cal M}_{0}(\Delta t)$
is associated with expressing the atom state
$|\Psi_A(t)\ra  = \alpha(t)|e\ra + \beta(t)|g\ra $
in terms of a column vector
$[\alpha(t),\beta(t)]^T$,
which makes well defined the action of
${\cal M}_{0}(\Delta t)$ on the atom state.

Taking the rectangular SDF shown in Fig.\ 2(c) as an example,
we display in Fig.\ 4 the results of quantum trajectory simulations,
based on the Monte-Carlo algorithm proposed above.
In Fig.\ 4(a), (b) and (c) we show, respectively,
a single representative trajectory for $x=0.02$, 0.2 and 2;
and for each case we indicate also the events of photon emissions.
We observe that, for smaller $x$ (smaller $\tau$),
the photo emissions are scarcer.
This is nothing but the consequence of the Zeno effect.
This phenomenon can be observed only for photon-detection
in finite-bandwidth non-Markovian environment.
For wide-band-limit Markovian environment, the results are
independent of $\tau$ \cite{SG13,WM09,Jac14,Dali92,WM93}.

In Fig.\ 4(d), for $x=0.2$ as an example,
we show the result of ensemble-average over 5000 trajectories
and compare it with the result from
an ensemble-averaged master equation, say,
the Lindblad master equation
with photon emission rate of $\gamma_{\rm eff}$.
We find perfect agreement between them.
We may remark that, the average result shown in Fig.\ 4(d)
is not the usual {\it reduced} dynamics
after tracing out the degrees of freedom of the environment,
despite that tracing means also averaging all the measurement results.
But the {\it average} associated with the {\it reduced} state
is done only at the concerned time instant $t$;
before $t$, there are no measurement interruptions
-- as involved in contrast by the quantum trajectory simulation --
to the entangled evolution of the coupled system-and-environment.

More detailed discussion about this issue
is referred to Ref.\ \cite{Xu16}
and references therein (the series of works by Wiseman {\it et al}).

\section{Discussion and Summary}

In the existing QT theory and simulations, the time step $\Delta t$ is assumed
such that during $\Delta t$ there is at most one photon emitted/detected,
while the (smaller) measurement time interval $\tau$ is implied
to account for the continuous (or, frequent) measurements.
To our knowledge, it has not been well clarified
that the existing QT result is free from the choice of $\tau$ \cite{SG13}.
This implies that, in the existing QT theory, the choice of the
measurement interval $\tau$ can be relaxed to $\Delta t$ \cite{SG13}.
However, our result, 
which is associated with finite bandwidth $\Lambda$
and small measurement time interval $\tau$,
shows that the result depends on the choice of $\tau$.
For instance, different choice of $\tau$ can result in big differences
such as dynamical phase transition in the counting statistics
of spontaneous emissions \cite{XL16}.

On the other aspect, if we let the interval $\tau=\Delta t$,
this defines the scaling parameter $x=\Lambda\,\Delta t$.
Owing to the scaling property,
for not very narrow $\Lambda$ and not long $\Delta t$,
we cannot distinguish the result from the one given by
different $\Lambda$ and $\tau$ but with the same $x$ ($x=\Lambda\tau$).
Only for very narrow bandwidth $\Lambda$ and relatively long $\Delta t$,
the result differs from that given by the same $x$
under choice of large $\Lambda$ and small $\tau$.
However, for any cases,
\Eq{til-alph-t} is a good starting point for the QT simulations.
For the former case, the non-Markovian memory effect is killed
by the frequent interruptions of measurement,
which reduce the solution of \Eq{til-alph-t} to
the simpler result of \Eq{at-mkv}, or Eqs.\ (\ref{abar-2}) and (\ref{gamma-x}).
For the latter case, the memory-involved iteration based on
\Eq{til-alph-t-2} can be implemented.

Generally speaking, the finite bandwidth environment will result in memory effect
in the reduced dynamics of the {\it system of interest}.
This is typically manifested by a time-convolutional form
of master equation for the reduced state,
just as observed also in \Eq{til-alph-t}.
In this sense, we call the environment with finite bandwidth a non-Markovian one.
However, in the presence of frequent measurement interruptions,
the non-Markovian memory effect is largely destroyed.
The main consequence of the non-Markovian (finite bandwidth) environment
is the dependence of the measurement interval $\tau$,
or more interestingly, of the scaling variable $x=\Lambda\tau$.

To summarize, we have generalized the measurement theory and the associated QT approach
to environment with finite bandwidth and beyond the Lorentzian spectrum.
For finite $\Lambda$ and small $\tau$, which results in a finite or small
parameter $x=\Lambda\tau$, the null-result conditioned state evolution
and quantum jump probability will be drastically affected
by the choice of $\tau$, or more interestingly, by the scaling parameter $x$.

For arbitrary SDF, we generally proved the existence of scaling property.
For Lorentzian and some non-Lorentzian SDFs,
by keeping $x=\Lambda\tau$ fixed but making the limits
$\Lambda\to\infty$ and $\tau\to 0$,
we obtained analytic result
to facilitate the QT simulation for finite $\Lambda$ and $\tau$
(but with the same $x$), owing to the underlying scaling property.

However, even if the analytical result is not available,
one can still use \Eq{til-alph-t-2} or Eqs.\ (\ref{abar-2}) and (\ref{gamma-x})
to simulate the quantum trajectories.

\vspace{0.3cm}
{\flushleft\it Acknowledgements.}---
This work was supported by the
National Key Research and Development Program of China
(No 2017YFA0303304), and the National Natural Science Foundation of China(No 11675016).


\appendix

\section{Connection with the KK Solution}

In this Appendix we make a connection of our treatment
leading to Eqs.\ (\ref{abar-2}) and (\ref{gamma-x})
with the Kofman-Kurizki (KK) solution of Eq.\ (12) in Ref.\ \cite{Kur00},
which was employed there to analyze the Zeno and anti-zeno effects.
Starting with \Eq{til-alph-t}, i.e.,
$\dot{a}(\tau)=-i\int^{\tau}_0 du \tilde{F}(u) a(\tau-u)$,
following Ref.\ \cite{Kur00}, in the limit of short $\tau$
we alternatively have
\begin{equation}\label{A1}
\dot{a}(\tau)\simeq -i\int^{\tau}_0 du \tilde{F}(u)  \,.
\end{equation}
Here the limit consideration
$\alpha(\tau-u)\rightarrow \alpha(0)=1$ in the integrand was inserted.
This consideration has a formal difference from the one
leading to \Eq{at-mkv} in the main part of this work.
Nevertheless, we will find both treatments equivalent.

Integrating the both sides of \Eq{A1} we obtain
\begin{eqnarray}
a(\tau)-a(0)
&=&-i\int^{\tau}_0 du\left[\int^{u}_0 du^\prime\tilde{F}(u^\prime)\right]  \nl
&=&-i\int^{\tau}_0 \mathrm{d}u \, (\tau-u)\tilde{F}(u)  \,.
\end{eqnarray}
In deriving the second equality we have used the technique of integration by parts.
Actually this is the KK solution Eq.\ (10) given in Ref.\ \cite{Kur00}.

Now let us consider $\bar{a}(t)= [a(\tau)]^n$
in the short-$\tau$-limit under the condition $t=n\tau$.
Noting that $a(0)=1$, we have
\begin{eqnarray}
[a(\tau)]^n&=&\left[1-i\int^\tau_0 du \, (\tau-u)\tilde{F}(u)\right]^n\nonumber\\
&\simeq& 1-i\, n\int^\tau_0 du\, (\tau-u)\tilde{F}(u)\nonumber\\
&=& 1-i\, t\int^\tau_0 du\, (1-u/\tau)\tilde{F}(u)\nonumber\\
&\simeq&  e^{-rt/2} \,.
\end{eqnarray}
In the result of the last line,
the following rate parameter was introduced
\begin{eqnarray}\label{r-x-scaling}
r = 2i \int_0^\tau du\, (1-u/\tau) \tilde{F}(u)  \,.
\end{eqnarray}
The real part of this quantity, say, $R=\mathrm{Re}(r)$,
gives rise to the result of Eq.\ (12) in Ref.\ \cite{Kur00}.
Notice that, in Ref.\ \cite{Kur00},
$P_e(t=n\tau)=|a(\tau)|^{2n}\simeq e^{-Rt}$
was interpreted as the {\it survival probability}
under frequent observations in the context of Zeno effect,
while in our present work $[a(\tau)]^n$ is inserted
into \Eq{null-Psi} as a {\it null-results conditioned change}
of the superposition amplitude which is in general a complex number.

Below we employ \Eq{r-x-scaling} to prove the {\it scaling property},
with the help of {\it the key structure
observed by \Eq{gx} and proved in general by \Eq{gx-prove}}.
As in the main part, let us introduce the scaling parameters
$x=\Lambda\tau$ and $x^\prime=\Lambda u$.
We then reexpress \Eq{r-x-scaling} as follows:
\begin{eqnarray}\label{r-x-2}
r&=& 2i\, \int_0^{x^\prime/\Lambda} d(x^\prime/\Lambda)
\, (1-x^\prime/x)\tilde{F}(x^\prime/\Lambda)\nonumber\\
&=&\frac{2i}{x}\int_0^x dx^\prime \,  (x-x^\prime) g(x^\prime)
\,\equiv\, r(x)   \,,
\end{eqnarray}
which shows that $r$ is a function of the joint variable
$x=\Lambda\tau$, i.e., holding the desired scaling property.
We should emphasize that, in order to achieve the proving,
using here the property $\tilde{F}(x^\prime/\Lambda)=\Lambda g(x')$
proved in \Eq{gx-prove} is a key point.

We now prove that this result is nothing but the one given by \Eq{gamma-x},
which is obtained in the main part by a different limiting treatment.
Starting with \Eq{gamma-x}, the proving simply reads:
\bea
\gamma(x)&=& \frac{2i}{x}\int_0^x dx^\prime\int_0^{x^\prime}
dx^{\prime\prime} \, g(x^{\prime\prime}) \nonumber  \\
&=& \frac{2i}{x}\left[x\int^{x}_0 \mathrm{d}x^{\prime\prime}
g(x^{\prime\prime}) -\int^x_0 x^\prime
g(x^\prime)\mathrm{d}x^\prime\right]\nonumber\\
&=& \frac{2i}{x}\left[x\int^{x}_0
\mathrm{d}x^\prime g(x^\prime) -\int^x_0 x^\prime g(x^\prime)
\mathrm{d}x^\prime\right]\nonumber\\
&=& \frac{2i}{x}\int^x_0 \mathrm{d}x^\prime \,(x-x^\prime)g(x^\prime) \,.
\eea
This is the result of $r(x)$ given by \Eq{r-x-2}.


\end{CJK*}
\end{document}